\documentclass[
    ,final            
  ]
  {aipproc}

\layoutstyle{8x11double}

\usepackage{amsmath}
\usepackage{amssymb}

\newcommand{\fig}{Fig.}
\newcommand{\ie}{{\it i.e.}}
\newcommand{\eg}{{\it e.g.}}
\newcommand{\etc}{{\it etc.}}
\newcommand{\Ref}{Ref.}
\newcommand{\Refs}{Refs.}
\newcommand{\Tab}{Table}

\newcommand{\ldm}{\Delta m_{31}^2}
\newcommand{\sdm}{\Delta m_{21}^2}
\newcommand{\deltacp}{\delta_{\mathrm{CP}}}
\newcommand{\stheta}{\sin^2 2 \theta_{13}}

\newcommand{\figu}[1]{\fig~\ref{fig:#1}}

\begin{document}

\title{On near detectors at a neutrino factory.}

\classification{14.60.Pq, 13.15.+g, 14.60.St
                }
\keywords      {Neutrino oscillations, neutrino factory, non-standard interactions}

\author{Jian Tang\thanks{Speaker email: jtang@physik.uni-wuerzburg.de}, 
Walter Winter\thanks{email: winter@physik.uni-wuerzburg.de}}{
  address={Institut f{\"u}r Theoretische Physik und Astrophysik, Universit{\"a}t W{\"u}rzburg, \\
       D-97074 W{\"u}rzburg, Germany}
}


\begin{abstract}
The geometric effects of the beam in near detectors at a neutrino factory are discussed. The refined systematics treatment, including cross section errors, flux errors and background uncertainties, is compared with the IDS-NF one. Different near detector setups are included. We also probe their effects both at the measurements of standard neutrino oscillation parameters and constraints of the non-standard neutrino interaction.
\end{abstract}

\maketitle


\section{Introduction}
The design of a neutrino factory has been put forward and discussed in international studies, such as in \Refs~\cite{Albright:2000xi, Apollonio:2002en,Albright:2004iw,Bandyopadhyay:2007kx}. Especially the most recent study, the International scoping study of a future Neutrino Factory and super-beam facility~\cite{Bandyopadhyay:2007kx,Abe:2007bi,Berg:2008xx}, has laid the foundations for the currently ongoing Design Study for the Neutrino Factory (IDS-NF)~\cite{ids}. This
initiative from about 2007 to 2012 is aiming to present a design report, schedule, cost estimate, and risk assessment for a neutrino factory. It defines a baseline setup of a high energy neutrino factory with two baselines
$L_1 \simeq 4 \, 000 \, \mathrm{km}$ and $L_2 \simeq 7 \, 500 \, \mathrm{km}$ (the ``magic'' baseline) operated by two racetrack-shaped storage rings, where the muon energy is 25~GeV (for optimization questions, see \Refs~\cite{Huber:2006wb,Gandhi:2006gu}). There are no Near Detector (ND) specifications yet, and the systematics treatment is done in an effective way by signal and background normalization errors uncorrelated among all channels and detectors. Therefore, there are a number of questions currently raised within the IDS-NF:
\begin{itemize}
\item
 Study of the potential of near detectors to cancel systematical errors.
\item
 Study of the characteristics of the near detectors, such as technology, number, \etc.
\item
 Study of the use of the near detectors for searches of new physics.
\end{itemize}
For the near detectors at a neutrino factory, a summary of options can be found in \Ref~\cite{Abe:2007bi}. In the work~\cite{Tang:2009na}, a more refined systematics treatment including cross section errors, flux errors, and background uncertainties is simulated to address several of these questions by GLoBES \Ref~\cite{Huber:2004ka,Huber:2007ji}. Both the physics results of the high energy neutrino factory for standard oscillations and also sensitivities of nonstandard interactions are presented.

\section{Definition of near detectors}

\begin{figure}[!thb]
\includegraphics[scale=1,width=0.9\textwidth]{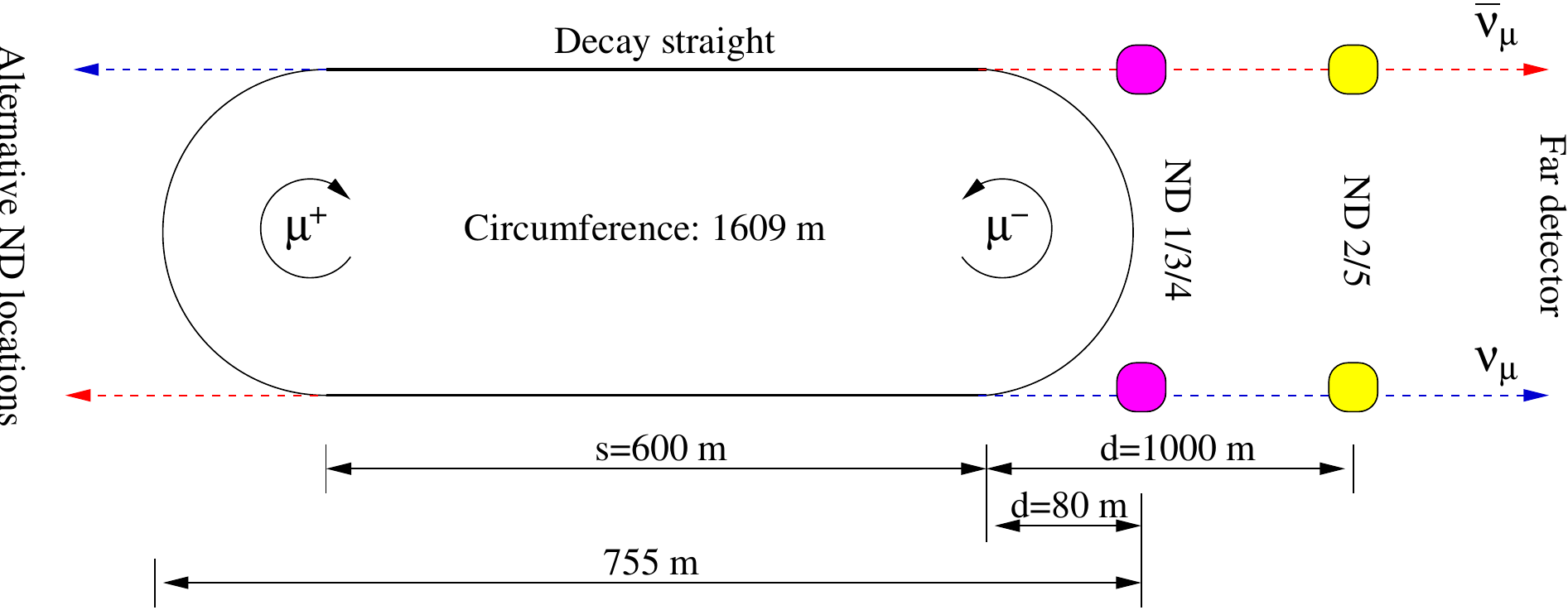}
\caption{\label{fig:ring}Geometry of the muon storage ring and possible near detector (ND) locations (not to scale). The baseline $L$ is the distance between production point and near detector, \ie, $d \le L \le d+s$. This figure is taken from \Ref~\cite{Tang:2009na}}
\end{figure}

At first, we consider the geometry of the storage ring and possbile near detector locations as depicted in Fig.~\ref{fig:ring} where $s$ is the length of the decay straight, $d$ is the distance from the nearest point in the decay straight to the detector and the baseline $L$ is the distance between the production point and the near detector. Then we could define limiting cases in terms of geometry.\\ 
\underline{Near detector limit.} In this case, the neutrino beam divergence is smaller than the detector diameter for the farthest decay point of the decay straight $L=d+s$, \ie, the full flux (integrated over the angle) is seen by the detector from any decay point in the straight.\\
\underline{Far detector limit.} In this case, the beam diameter given by the opening angle applied to the size of detector is of the order of the detector diameter at the nearest decay point $L=d$, where the near detector takes effects as a far detector for any decay point in the straight.

\begin{table}[b]
\begin{footnotesize}
\begin{tabular}{lrrrrr}
\hline 
Parameter & ND1 & ND2 & ND3 & ND4 & ND5   \\
\hline
Diameter $D$ & 17~m & 4~m & 4~m & 0.32~m & 6.8~m  \\
Distance $d$ & 80~m & 1000~m & 80~m & 80~m &  1000~m \\
Mass & 450~t & 25~t & 25~t & 0.2~t & 2000~t \\
\hline
\end{tabular}
\end{footnotesize}
\caption{\label{tab:nd} 
Definition of near detectors.
}
\end{table}

Our near detector parameters are shown in \Tab~\ref{tab:nd}. The fiducial volumes in terms of diameter, distance $d$ to the end of the decay straight, and mass. The fiducial volumes are assumed to be cylindrical. If the density is about $1 \, \mathrm{g \cdot cm^{-3}}$ (such as for a liquid scintillator), the active detectors will be about 2~m long for ND1 to ND4. There we define a hypothetical ND1 as a detector operating in the near detector limit, and a ND2 as a detector operating close to the far detector limit. ND3 is an intermediate case between the near and far limits, as we will demonstrate later.
The size of ND2 and ND3 is similar to conventional near detectors, such as SciBOONE, MINER$\nu$A, NOMAD or the MINOS near detector. ND4 is a smaller version of ND2 with the same ratio between detector diameter and distance $d$. If the source was a point source (\ie, the straight would be a point), the event rate would be exactly the same as in ND2. Finally, ND5 is an OPERA-like near detector, which we will only use for non-standard physics tests. It has to be mentioned that the impact of near detectors are limited by the statistics in the far detector. Further investigations also prove it by the fact that there is no significant difference for measurements of standard neutrino oscillation parameters from ND1 to ND4 so that we only refer to the ``near detector'' in the following to simplify the descriptions.

\section{Systematics treatment}
\begin{figure}[t]
\includegraphics[scale=1,width=0.9\textwidth]{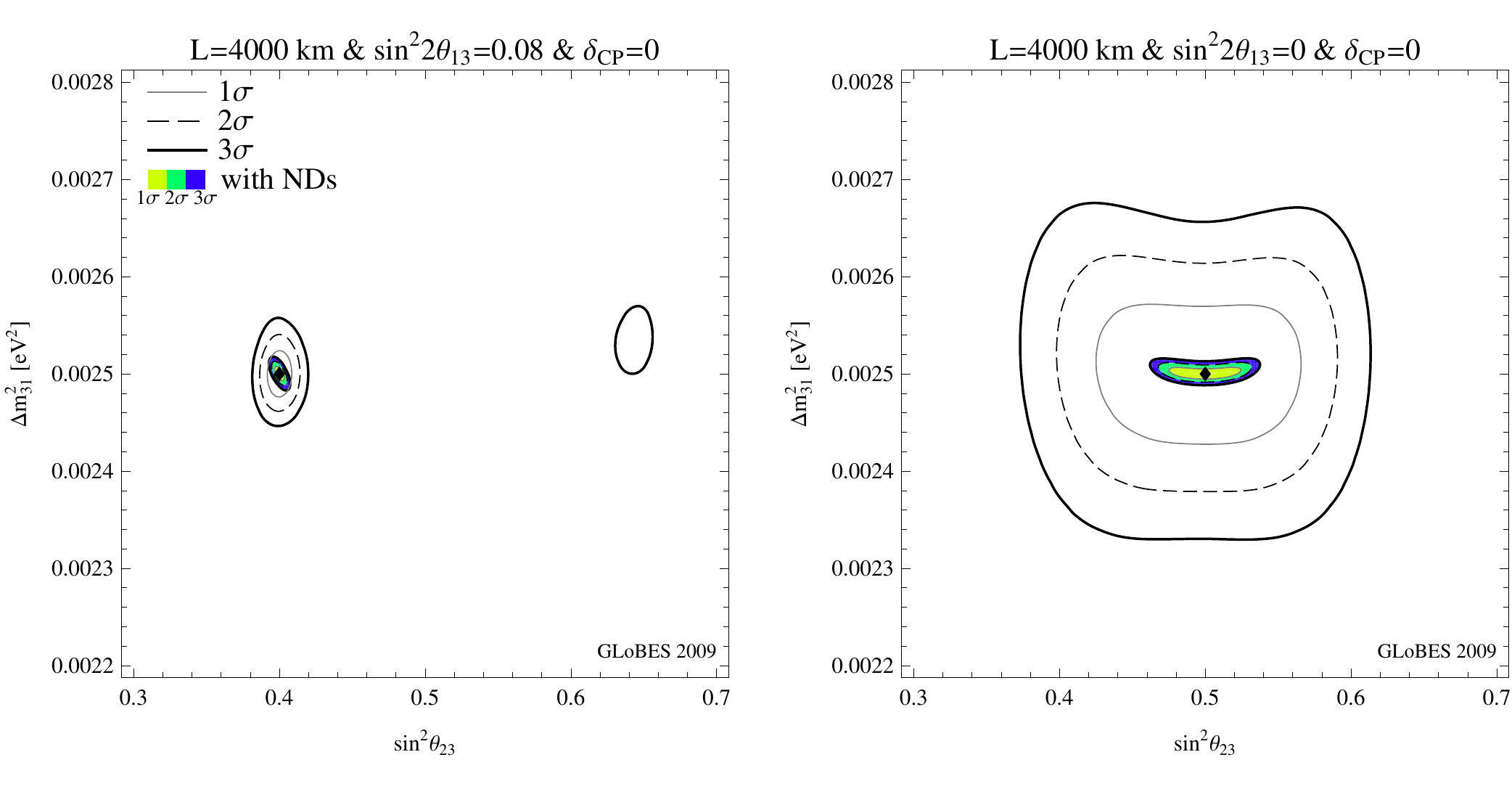}
\caption{\label{fig:atmhigh} The $\sin^2 \theta_{23}$-$\ldm$ allowed region  for a high energy neutrino factory at the $L=4000 \, \mathrm{km}$ baseline only; $1\sigma$, $2 \sigma$, $3\sigma$ CL (2 d.o.f.), best-fit points marked by diamonds. The filled contours correspond to our near detector-far detector simulation, whereas the unfilled contours represent the far detector only.
Normal hierarchy only, \ie, no sign-degenerate solution shown. The figures are similar to those in \Ref~\cite{Tang:2009na}.}
\end{figure}
In the following, we give the refined systematics treatment and compare it with the systematics treatment in the IDS-NF baseline setup.\\
\textbf{IDS-NF baseline setup:}\\
The currently discussed setup for a high energy neutrino factory is the IDS-NF baseline setup~\cite{ids}. It consists of two baselines $L_1 \simeq 4 \, 000 \, \mathrm{km}$
and $L_2 \simeq 7 \, 500 \, \mathrm{km}$, operated by two racetrack-shaped storage rings
simultaneously with both polarities ($\mu^+$ and $\mu^-$ stored, circulating
in different directions). The geometry of a storage ring is shown in \figu{ring}. There are no near detector specifications yet.
We define the polarities of the muons stored in the rings as:
\begin{eqnarray}
+ & : & \mu^- \rightarrow e^- + \bar\nu_e + \nu_\mu \\
- & : & \mu^+ \rightarrow e^+ + \nu_e + \bar\nu_\mu 
\end{eqnarray}
The following oscillation channels are used (with the corresponding muon polarities):
\begin{eqnarray}
\nu_\mu \texttt{appearance }(-) & : & \nu_e \rightarrow \nu_\mu \label{equ:numuapp} \\
\bar\nu_\mu \texttt{appearance }(+) & : & \bar\nu_e \rightarrow \bar\nu_\mu \label{equ:numubarapp} \\
\nu_\mu \texttt{disappearance }(+) & : & \nu_\mu \rightarrow \nu_\mu \label{equ:numudisapp} \\
\bar\nu_\mu \texttt{disappearance }(-) & : & \bar\nu_\mu \rightarrow \bar\nu_\mu \label{equ:numubardisapp}  
\end{eqnarray}
For the backgrounds, neutral currents are included for all channels, and mis-identified charged current events are included for the appearance channels.
Since there are two racetrack-shaped storage rings $S1$ and $S2$ targeted towards two far detectors, there are altogether
eight oscillation channels.
In the IDS-NF baseline setup~1.0, the systematics treatment is rather straightforward.  For each channel and baseline, 
an overall normalization error is included, which is $2.5\%$ for the signal rates, and $20\%$ for the background rates. The normalization errors are uncorrelated between signal and background, among different channels, detectors, and polarities, but fully correlated among all bins. The energy resolution is assumed to be $\Delta E \, \mathrm{[GeV]} = 0.55 \sqrt{E \, \mathrm{[GeV]}}$.\\
\textbf{Refined systematics treatment:}\\
Compared to the IDS-NF systematics, the different signal errors are correlated in a particular way. For instance, the cross section errors at the two far detectors are fully correlated, which will turn out to have interesting effects.\\
\underline{Flux normalization errors} are fully uncorrelated among the different polarities $+$, $-$ and storage rings $S1$, $S2$, but fully correlated among all bins and all channels operated with the same beam.\\
\underline{Cross section errors} for the inclusive charged current cross sections are fully correlated among all signal and background channels measuring $\nu_\mu$ or $\bar\nu_\mu$, but fully uncorrelated among all bins.\\
\underline{Background normalization errors} are fully correlated among all bins, but fully uncorrelated among all channels, polarities, and detectors.

\section{Simulation results}
\begin{figure}[!thb]
\includegraphics[scale=1,width=0.9\textwidth]{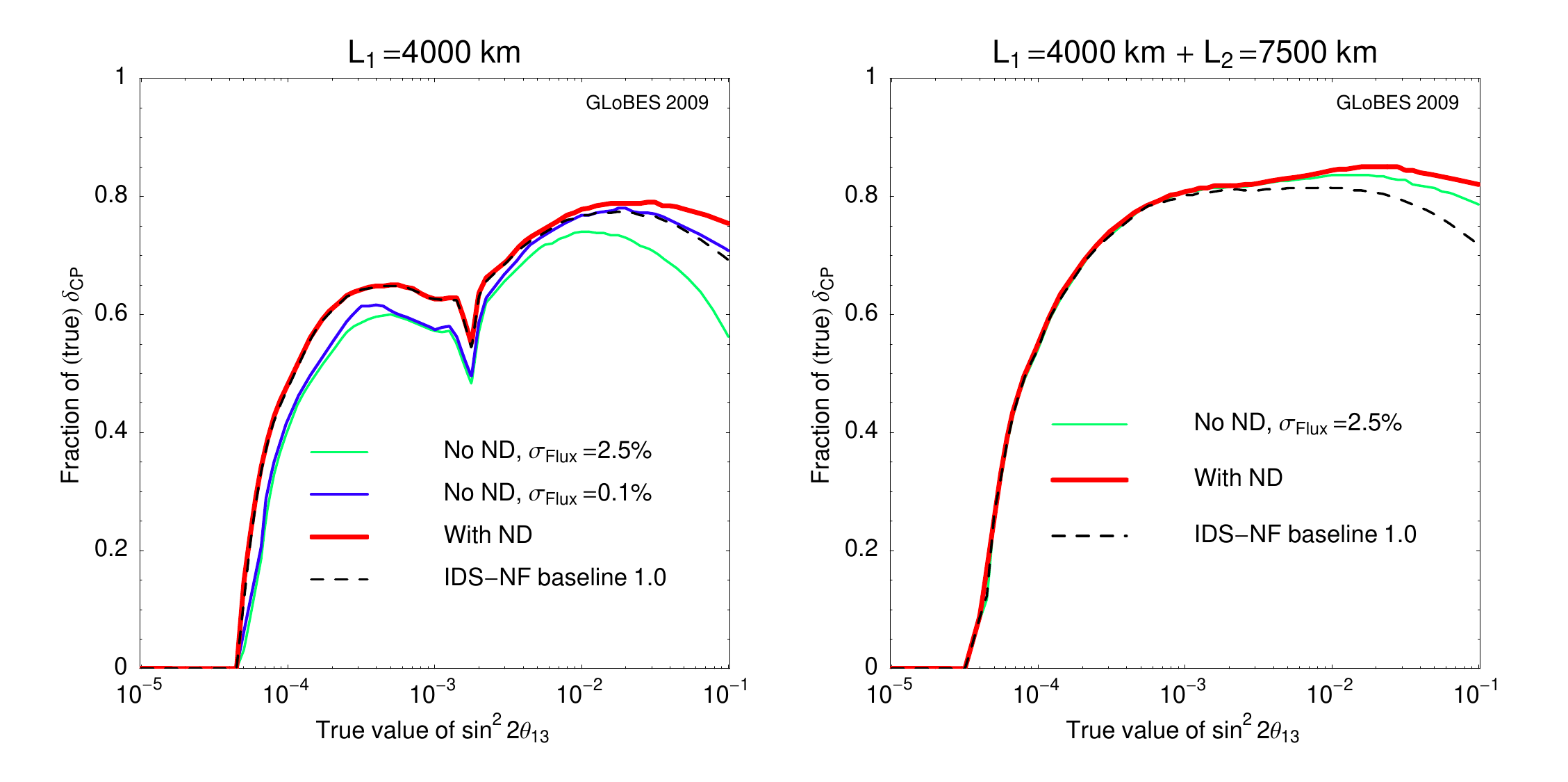}
\caption{\label{fig:ndcomp} CP violation discovery reach as a function of true $\stheta$ and the fraction of (true) $\deltacp$ for one far detector (left) and two far detectors (right) at $3\sigma$ CL. The figures are taken from \Ref~\cite{Tang:2009na}.}
\end{figure}
For the experiment simulation, we use the GLoBES software~\cite{Huber:2004ka,Huber:2007ji} with user-defined systematics. 
For the oscillation parameters, we use (see, \eg, \Refs~\cite{GonzalezGarcia:2007ib,Schwetz:2008er}) $\sin^2 \theta_{12}=0.3$, $\sin^2 \theta_{23} = 0.5$, $\sdm = 8.0 \cdot 10^{-5} \, \mathrm{eV^2}$, $\ldm = 2.5 \cdot 10^{-3} \, \mathrm{eV^2}$, and a normal mass hierarchy, unless specified otherwise. We impose external errors on $\sdm$ and $\theta_{12}$ of 4\% each, and we include a 2\% matter density uncertainty~\cite{Geller:2001ix,Ohlsson:2003ip}.\\
\textbf{Standard oscillation measurements:}\\
In Fig.~\ref{fig:atmhigh}, we show $\sin^2 \theta_{23}$-$\ldm$ allowed region  for a neutrino factory at the $L=4000 \, \mathrm{km}$ baseline only. The impact of near detectors is illustrated with filled contours. In the right panel, the effect of the near detectors is very large for maximal mixing. Meanwhile, the effect in the left panel is less dramatic than in the right one but still substantial, since the filled contours are limited by the statistics in the far detector. Especially, the octant degeneracy can be excluded with the near detectors at a high confidence level (if $\stheta$ is large enough). 

In Fig.~\ref{fig:ndcomp}, we show the CP violation discovery reach as a function of the true $\sin^22\theta_{13}$ and the fraction of (true) $\delta_{CP}$ for one far detector in the left panel and two detectors in the right panel. For the $\theta_{13}$ and CP violation discovery reaches, we have not found any significant impact, because these measurements are background limited. For the CP violation discovery reach as depicted in Fig.~\ref{fig:ndcomp}, the near detectors are important in the left panel, because the unknown atmospheric parameters lead to intrinsic unknown backgrounds from the CP-even terms, which can be controlled by the near detectors.
For a two baseline high energy neutrino factory, such as the IDS-NF baseline setup with $L_1=4000 \, \mathrm{km}$ combined with $L_2=7500 \, \mathrm{km}$, we have demonstrated that the considered systematical errors cancel even without near detectors and good flux monitoring. In such a setup, the same product of fluxes and cross sections is measured in both far detectors. Therefore, a two baseline neutrino factory should be very robust with respect to systematics, no matter how many near detectors are used.\\
\textbf{Non-Standard interaction sensitivites:}\\
Non-standard interactions (NSI)~\cite{Wolfenstein:1977ue,Valle:1987gv,Guzzo:1991hi,Grossman:1995wx,Roulet:1991sm}  are effects of physics beyond the Standard Model. Near detectors may be very relevant for the extraction of the source NSI, since these lead to an unambiguous ``zero-distance'' ($L \ll L^{\mathrm{osc}}$) effect proportional to $|\epsilon^s|^2$~\cite{Gonzalez-Garcia:2001mp, Huber:2002bi}.

\begin{table}[b]
\begin{tabular}{lrrr}
\hline
& Without ND5 & & With ND5 \\
\hline
$|\epsilon^s_{e \tau}|$ & 0.004 & & 0.0007\\
$|\epsilon^s_{\mu \tau}|$ & 0.4 & & 0.0006 \\
$|\epsilon^m_{e \tau}|$ & 0.004 & & 0.004 \\
$|\epsilon^m_{\mu \tau}|$ & 0.02 & & 0.02 \\
\hline
\end{tabular}
\caption{\label{tab:nsi} Sensitivity to the non-diagonal NSI parameters where the phases have been marginalized over.}
\end{table}

Because there are no tau neutrinos in the beam, the most interesting option may be to use near detectors to measure $\nu_\tau$ appearance, see, \eg, \Refs~\cite{Antusch:2006vwa,FernandezMartinez:2007ms,Malinsky:2008qn,Malinsky:2009gw}.  Therefore, we use the high energy neutrino factory with ND3 and two (symmetrically operated) additional 2~kt OPERA-like emulsion cloud chambers (ECC) as simulated in \Ref~\cite{Huber:2006wb} and described in \Ref~\cite{Autiero:2003fu} for the $\nu_e \rightarrow \nu_\tau$ channel at the neutrino factory, and operate them at $L=1 \, \mathrm{km}$ corresponding to ND5 (see also \figu{ring}). The NOMAD limits for the zero-distance effect could be improved by about two orders of magnitude. In addition, if it is possibe that source and matter NSI are connected at a generic level assuming dimension six effective operators generated at the tree level by messagers from a high energy scale~\cite{Gavela:2008ra}, even the matter NSI parameter $|\epsilon^m_{\mu \tau}|$ becomes quite strongly limited, exceeding the bound from lepton universality. Furthermore, CP violation from matter NSI may become measurable down to 0.0005 in $|\epsilon^m_{\mu \tau}|$.

\section{Conclusions}
We have included the geometry of the source and the detectors in the event rate calculations in the definitions of qualitatively different near detectors. In addition, we have refined the systematics treatment including cross section errors, flux normalization errors and background uncertainties. For the characteristics of the near detectors for standard oscillation measurements, we have found that two near detectors should be operated on-axis (with respect to the decay straights) if the muons and anti-muons circulate in different directions in a racetrack-shaped ring. If the neutrino factory has only one baseline, such as a high energy ($E_\mu=25 \, \mathrm{GeV}$) neutrino factory with $L=4000 \, \mathrm{km}$, the near detectors are mandatory for the leading atmospheric parameter measurements.
For the $\theta_{13}$ and mass hierarchy discovery reaches, we have not found any significant impact, because these measurements are background limited. For the CP violation discovery reach, the near detectors are important, because the unknown atmospheric parameters lead to intrinsic unknown backgrounds from the CP-even terms, which can be controlled by the near detectors. For a two baseline high energy neutrino factory, such as the IDS-NF baseline setup with $L_1=4000 \, \mathrm{km}$ combined with $L_2=7500 \, \mathrm{km}$, we have demonstrated that the considered systematical errors cancel even without near detectors and good flux monitoring.
Finally, we have considered OPERA-like near detectors for $\nu_\tau$ appearance to measure non-standard interactions (NSI).  We have demonstrated the NOMAD limits for the zero-distance effect could be improved by about two orders of magnitude. In a word, two near detectors are mandatory for a successful neutrino factory operation with one baseline, and a good enough flux monitoring will be useful for some physics measurements. However, a two baseline high energy neutrino factory may prove to be very robust with respect to systematical errors, even if the systematics goals in the initial one baseline operation phase cannot be achieved.

\begin{theacknowledgments}
JT appreciates the scholarship and good organizations in NuFact09.
JT is supported to attend the conference by Deutsche Forschungsgemeinschaft through the Research Training Group
GRK 1147 Theoretical Astrophysics and Particle Physics.

\end{theacknowledgments}

\bibliographystyle{aipproc}


\end{document}